\documentclass[twocolumn]{aastex701}
\usepackage{amsmath}
\usepackage[utf8]{inputenc}

\newcommand{\dif}{\mathrm{d}}
\newcommand{\e}{\mathrm{e}}
\newcommand{\yr}{\,\mathrm{yr}}
\newcommand{\Myr}{\,\mathrm{Myr}}
\newcommand{\Gyr}{\,\mathrm{Gyr}}
\newcommand{\AU}{\,\mathrm{AU}}

\newcommand{\ME}{M_{\oplus}}
\newcommand{\RE}{R_{\oplus}}

\newcommand{\MSol}{M_{\odot}}

\newcommand{\affilciera}{\affiliation{Center for Interdisciplinary Exploration and Research in Astrophysics (CIERA), Northwestern University, 1800 Sherman Ave., Evanston, IL 60201, USA}}
\newcommand{\affilnupa}{\affiliation{Department of Physics and Astronomy, Northwestern University, 2145 Sheridan Rd., Evanston, IL 60208, USA}}

\shorttitle{Breaking the chains with planetesimals}
\shortauthors{Li, O'Connor \& Rasio}

\begin{document}

\title{Intruder Alert: Breaking Resonant Chains with Planetesimal Flybys}

\correspondingauthor{Jiaru Li ({\tt jiaru.li@northwestern.edu}),\\Chris O'Connor ({\tt christopher.oconnor@northwestern.edu})}

\author[0000-0001-5550-7421,sname=Li,gname=Jiaru]{Jiaru Li}
\altaffiliation{Joint lead authors}
\affilciera
\email{jiaru.li@northwestern.edu}

\author[0000-0003-3987-3776,sname='O\'Connor',gname='Christopher E.']{Christopher E.\ O'Connor}
\altaffiliation{Joint lead authors}
\affilciera
\email{christopher.oconnor@northwestern.edu}

\author[sname=Rasio,gname='Frederic A.']{Frederic A.\ Rasio}
\affilciera
\affilnupa
\email{rasio@northwestern.edu}

\submitjournal{ApJ Letters}

\begin{abstract}
    The orbital architectures of compact exoplanet systems record their complicated dynamical histories. 
    Recent research supports the ``breaking-the-chains'' hypothesis, which proposes that compact systems 
    typically form in chains of mean-motion resonances (MMRs) but subsequently break out on a $\sim 100 \Myr$ timescale. 
    We investigate a scenario for breaking the chains through intermittent flybys of planetesimals originating from a distant reservoir. 
    Using $N$-body simulations and semi-analytical calculations, we characterize the disruption of MMRs through these flybys. 
    We find a planetesimal reservoir of total mass $\gtrsim 0.04 \ME$ is required to disrupt MMR chains, depending on the mass distribution and the typical number of flybys executed by each planetesimal. 
    We verify that systems disrupted in this way are frequently unstable to close encounters within $\sim 100 \Myr$ of the final flyby. 
    This mechanism operates in systems with both a sufficiently massive reservoir and an efficient mechanism for planetesimal injection. 
    Consequently, we predict an anti-correlation between resonant inner systems and dynamically active outer configurations. 
\end{abstract}


\section{Introduction}

Compact multi-planet systems, also known as ``{\it Kepler}-multis'' (KMs), 
constitute the majority of known planetary systems in the Galaxy.
These systems typically contain $\sim 3 \mbox{--} 6$ planets between $1$ and $4 \RE$ in size (``super-Earths'' or ``sub-Neptunes'') with orbital periods $\lesssim 400 \, {\rm d}$ \citep[e.g.,][]{Fressin+2013, Zhu+2018}. 
Extensive research over the past 15 years has attempted to relate the demographics of KM systems 
to the major processes governing their formation and dynamical evolution. 
Mean-motion resonances (MMRs) may be a particularly important aspect of the dynamical evolution of KMs. 
Although most KM systems are not observed to be in resonance today, 
the distribution of period ratios among adjacent planet pairs contains distinct features near low-order commensurabilities. 
A minority of systems comprise chains of multiple adjacent planets in resonance, 
such as Kepler-223 \citep{Mills2016_Kepler223chain}, TRAPPIST-1 \citep{Gillon2017_TRAPPIST1, Luger2017_Trappist1chain}, 
TOI-178 \citep{Leleu2021_TOI178chain}, and HD\,110067 \citep{Luque2023_HD110067chain}. 
Such architectures are the expected outcome of convergent type-I migration of planetary cores in the protoplanetary disk \citep[e.g.,][]{GoldreichTremaine1980,TerquemPapaloizou2007,WongLee2024}. 

Many features of the KM population can be reproduced under the assumption that compact super-Earth systems predominantly form in low-order MMR chains 
and, in most cases, subsequently experience a giant-impact phase due to dynamical instabilities 
\citep[e.g.,][]{Izidoro2017_breakingchains, Izidoro2021_breakingchains, Goldberg2022_architectures, GhoshChatterjee2024, Liveoak2024_closeneptunes, LiRixin2025_brokenchains}. 
This ``breaking-the-chains'' hypothesis between the physical properties and orbital architectures of KM systems. 
For example, planets in resonance have larger gaseous envelopes than those outside of resonance but are nonetheless less massive on average \citep{Leleu2024_resonantpuffs, Dai2024_chains, LiRixin2025_brokenchains}. 
This would be consistent with previous giant impacts in the non-resonant systems, 
which roughly doubled the planets' core masses through mergers while partially or completely eroding their envelopes. 

Further evidence for breaking-the-chains comes from recent age-resolved studies of planetary architectures. 
\citet{HamerSchlaufman2024_youngchains} compared the Galactic velocity distributions 
of the host stars of resonant and non-resonant KM systems. 
They found the resonant sample is kinematically cooler than the non-resonant, indicating that stars in the former are relatively young. 
\citet{Dai2024_chains} subsequently examined a sample of compact planetary systems with well-determined ages, 
for the first time robustly separating ``young'' systems (ages $< 100 \Myr$) from ``adolescent'' ($100 \Myr \mbox{--} 1 \Gyr$) and ``mature'' ($> 1 \Gyr$) systems. 
They reported that the fraction of systems with at least one pair of adjacent planets near a first- or second-order commensurability) 
decreases with age, 
from $(86 \pm 13) \%$ for young systems to $(38 \pm 12) \%$ and $(23 \pm 3)\%$ for adolescent and mature systems, respectively. 
This associates the dissolution of resonant chains with a characteristic timescale of $\sim 100 \Myr$ -- a key clue for dynamical studies seeking an explanation. 
 
Low-mass resonant chains formed through convergent migration in simple disk models tend to be long-term stable 
(e.g., \citealt{Izidoro2017_breakingchains}, \citealt{LammersWinn2024_HD110067}; 
but see \citealt{Pichierri2020_chainstability} and \citealt{Goldberg2022_chainstability}). 
The occurrence of intricate resonant chains in a subset of mature systems, such as those listed above, attests to this. 
Hence, breaking the chains requires one or more additional physical processes 
to trigger instability in the majority of systems on a $\sim 100 \Myr$ timescale. 
Previous studies have examined how a variety of dynamical and evolutionary processes affect KM architectures, 
including resonant repulsion via eccentricity damping \citep{LithwickWu2012_repulsion, Goldreich2014_overstable, Deck2015_mmrcapture, Nesvorny2022_mmrmigration}, 
mass changes through photoevaporation or stellar winds \citep{MatsumotoOgihara2020_breakingchains}, 
sweeping secular resonances 
\citep{Spalding2018_keplerobliquity, SpaldingMillholland2020_keplerobliquity, Faridani2023_sweeping, Faridani2025_sweeping}, 
and perturbations from distant giant planets 
\citep[e.g.,][]{Hansen2017_giantplanets, Huang+2017, PuLai2018_cjsecular, PuLai2021_cjscattering, RodetLai2021, BitschIzidoro2023_bullies, SobskiMillholland2023_cjsculpting, Sandhaus+2025}. 

Planet formation is a messy, inefficient process, leaving behind numerous residual planetesimals never accreted onto planetary cores. 
In the Solar System, a small fraction of these objects survive as asteroids, Kuiper Belt objects, and comets. 
Gravitational interactions between the residual planetesimals and the planets 
were likely a major driver of the Solar System's early dynamical evolution \citep[e.g.,][]{FernandezIp1984, Malhotra1995, Tsiganis+2005_Nice, Gomes+2005}. 
As compact exoplanet systems presumably host their own planetesimal populations, 
several previous studies have addressed how planetesimal scattering affects 
the evolution and stability of compact exoplanet systems near resonance 
\citep[e.g.,][]{Moore2013_planetesimalsKOI730, ChatterjeeFord2015_planetesimal, Raymond2022_trappist1planetesimals, GhoshChatterjee2023_planetesimals, HamerSchlaufman2024_youngchains, Wu2024_pingpong}. 

\begin{figure*}[th!]
    \centering
    \includegraphics[width=0.9\textwidth]{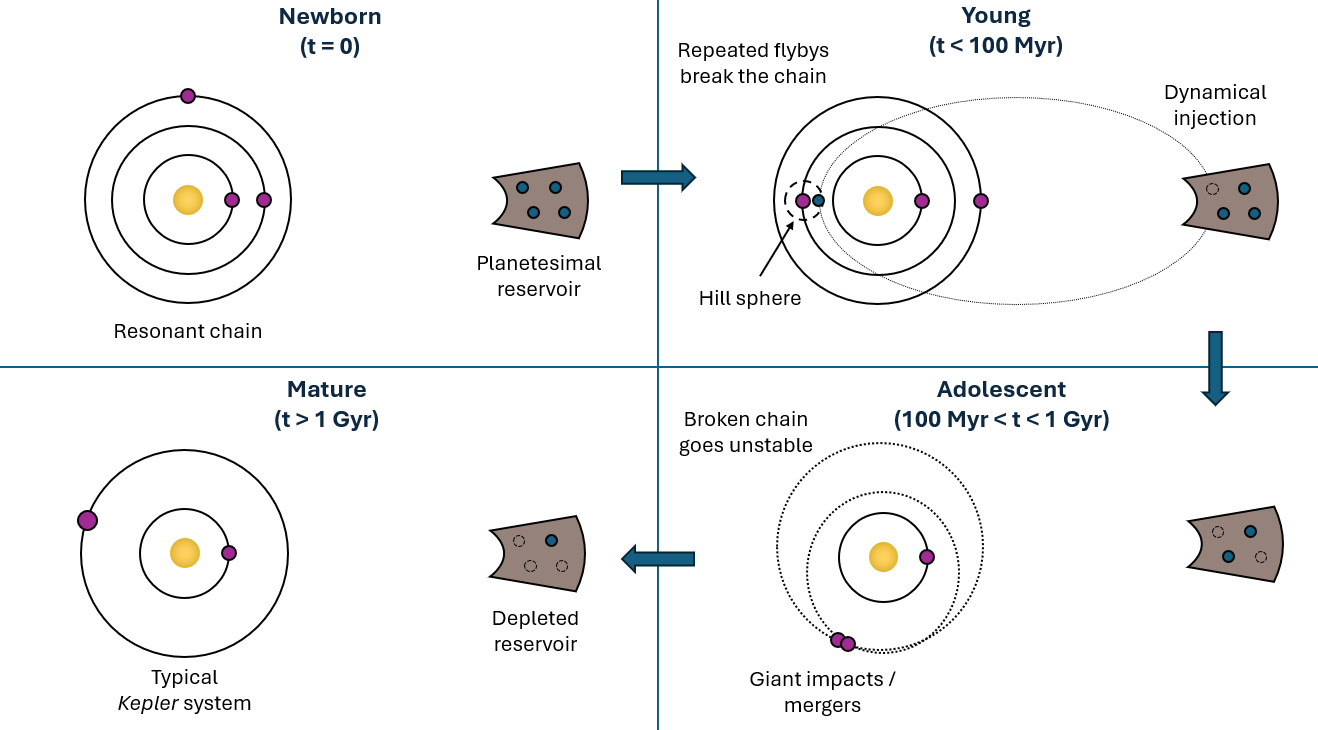}
    \caption{Cartoon illustration of the breaking of compact resonant chains through planetesimal scattering. 
    The arrow of time proceeds clockwise, starting from the upper left.}
    \label{fig:cartoon}
\end{figure*}

In this Letter, we describe the breaking of resonant chains in KM systems via planetesimal flybys. 
We study a situation where planetesimals are intermittently injected on highly eccentric orbits into a compact resonant chain, 
gradually breaking the planets out of resonance through weak gravitational scattering over a series of flybys. 
Figure \ref{fig:cartoon} illustrates our scenario in cartoon form. 
A key difference aspect our work is that we consider planetesimals delivered gradually from an outlying reservoir, 
rather than interspersed among the planets at birth. 
We thus relate the architectures of KM systems to the dynamical evolution of more distant bodies, 
recalling the intricate relationships among the terrestrial planets, Jovian planets, and minor bodies of the Solar System.

The Letter is organized as follows. 
In Section \ref{sec:analytical-model}, we describe the essence of resonance escape through planetesimal flybys with a simple analytical model predicting key scaling laws.
In Section \ref{sec:scattering_expts}, we use numerical scattering experiments to refine our model and quantify the probability of escaping resonance. 
In Section \ref{sec:instability}, we confirm the destabilization of compact planetary systems by the cumulative effect of flybys with further simulations. 
In Section \ref{sec:discussion}, we set forth necessary and sufficient conditions for our scenario, 
describe some caveats and possible extensions of the results, and summarize our main findings. 

\section{Theoretical motivation}
\label{sec:analytical-model}

Here we present a phenomenological model for escape from MMRs via repeated flybys based on the classic pendulum framework. 
As shown in Section \ref{sec:scattering_expts}, our predictions broadly agree with numerical experiments.

\subsection{Pendulum model for MMR}

The dynamics of a pair of planets in MMR is physically analogous to a simple pendulum, 
where the resonant argument $\theta$ corresponds to the angle of deflection from the vertical (up to an additive constant). 
Librating and circulating motions of a pendulum mirror the planets' dynamical behavior in and out of resonance, respectively. 
By extension, a chain of two-body MMRs is analogous to a system of coupled pendulums, one for each resonant argument: 
for small libration amplitudes, the collective motion can be described in terms of a set of normal modes of fixed amplitude; 
in this regime, the system's evolution is quasi-periodic and hence long-term stability is expected. 
On the other hand, if the mode amplitudes are sufficiently large, or if some modes marginally circulate, 
non-linear coupling between modes promotes chaos, potentially driving the system towards instability. 

The pendulum model is best understood by treating the inner planet ($l=1$) as a test particle near the $j:(j-k)$ commensurability. 
The relevant resonant argument is $\theta = (k-j) \lambda_{1} +j \lambda_{2} - k \varpi_{1}$, 
where $\lambda_{2} = n_{2} t$ with $t$ the time and $n_{2}$ the outer planet's mean motion. 
The Hamiltonian for the inner planet's orbital evolution is \citep[e.g.,][]{HaddenLithwick2018}
\begin{equation}
    \mathcal{H} = \frac{G M m_{1}}{a_{1}} \left[ - \frac{3}{4} (j-k)^{2} I^{2} + 2 \alpha \mu_{2} S_{j,k}(\alpha, e_{1}) \cos{\theta} \right],
\end{equation}
where $I = 2 (k-j) [(a_{1}/a_{1{\rm r}})^{1/2} - 1]$, a measure of the deviation from exact commensurability, 
is a dimensionless action conjugate to $\theta$. 
Its $x = \Delta a / a_{\rm i} \simeq $

We denote $\mu_{2} = m_{2} / M$ and $\alpha = a_{1}/a_{2} \simeq [(j-k)/j]^{2/3}$. 
The function $S_{j,k}(\alpha, e_{1})$ is taken to be a constant; 
to leading order in $e_{1}$, $S_{j,k} \simeq f_{d}(\alpha) e_{1}^{k}$, 
where $f_{d}(\alpha)$ is a combination of Laplace coefficients and their derivatives \citep{MurrayDermott1999}. 
Note $S_{j,k}$ may be positive or negative depending on $k$.
An analogous Hamiltonian describes the action of the inner planet on the outer, 
with slightly different coefficients and $\varpi_{2}$ replacing $\varpi_{1}$ in the expression for $\theta$. 

The equations of motion derived from $\mathcal{H}$ above provide a good approximation of resonant dynamics in general. 
A partial exception is for first-order resonances ($k=1$), 
for which this model is accurate only for sufficiently large eccentricities:
\begin{equation}
    e_{1} \gtrsim \left( \frac{\mu_{2}}{j} \right)^{1/3} \approx 0.017 \left( \frac{\mu_{2}}{10^{-5}} \frac{2}{j} \right)^{1/3}. 
\end{equation} 
This condition is violated in many of the systems we study, 
but the overall argument of this section stands even so. 

Since the pendulum Hamiltonian is integrable, 
the value of the Hamiltonian given by the initial conditions, $\mathcal{H}_{0} = \mathcal{H}(I_{0}, \theta_{0})$, 
is an integral of the motion describing the system's ``energy level.'' 
The libration amplitude can be expressed as the half-width of the level curve $\mathcal{H}(I, \theta) = \mathcal{H}_{0}$ along the $I$ dimension, 
given by the positive root of $\mathcal{H}(I_{\rm max}, \pi) = \mathcal{H}_{0}$. 
This, in turn, is related to the system's maximum deviation from exact commensurability: 
\begin{equation}
    x_{\rm max} \equiv \frac{\max|a_{1}(t) - a_{\rm 1r}|}{a_{\rm 1r}} = (j-k) I_{\rm max}.
\end{equation}

For a given $e_{1} \ll 1$, librating trajectories exist only within a narrow ``resonant bandpass'' given by $|I| \leq \Delta I_{\rm lib}$, 
where
$\Delta I_{\rm lib}$ is the half-width of the separatrix (i.e., the level curve passing through the unstable fixed point):
\begin{subequations} \label{eq:DeltaIlib}
\begin{align}
    \Delta I_{\rm lib} &\equiv \left( \frac{16}{3} \frac{\mu_{2} \alpha |f_{d}(\alpha)| e_{1}^{k}}{(j-k)^{2}} \right)^{1/2} \\
    &= 5 \times 10^{-4} \left( \frac{j-k}{2} \right)^{-1} \left( \frac{\mu_{2}}{10^{-5}} \frac{\alpha |f_{d}(\alpha)| e_{1}^{k}}{0.02} \right)^{1/2}
\end{align}
We denote the corresponding $x_{\rm lib} = (j-k) \Delta I_{\rm lib}$. 
\end{subequations}
For first-order resonances at low eccentricity, we require a slightly more complicated expression:
\begin{equation}
    \Delta I_{\rm lib}^{(k=1)} = \Delta I_{\rm lib} \left( 1 + \frac{\mu_{2} \alpha |f_{d}(\alpha)|}{27 (j-1)^{2} e_{1}^{3}} \right)^{1/2}, 
\end{equation}
where $\Delta I_{\rm lib}$ on the right-hand side is given by equation (\ref{eq:DeltaIlib}).

\subsection{Effect of repeated flybys}

Bearing the pendulum analogy in mind, one can develop intuition for why repeated planetesimal flybys break a system out of resonance.  
A single flyby has the same effect on a resonant as gently ``flicking'' a pendulum: 
that is, a small, abrupt change in the amplitudes and phases of the underlying normal modes. 
Successive flybys are statistically independent events in so far as the values of $\theta$ at any two flybys are uncorrelated;  
thus, the collective effect of many flybys occurring at random intervals should be diffusive evolution of the mode amplitude(s). 
The system escapes resonance when a mode amplitude crosses the threshold for circulation, i.e.\ when the planet's orbit has drifted by $|x| \gtrsim x_{\rm lib}$. 
In general, the modes involving the outermost planet in a chain exhibit the fastest diffusion 
because that planet presents the greatest scattering cross-section to incoming planetesimals. 


These considerations suggest we characterize the evolution of resonant planets subjected to repeated flybys as follows. 
We suppose one planet in the pair (usually the outer) scatters many planetesimals with a typical mass $m_{p} \ll m_{\rm out}$ 
and consequently undergoes a one-dimensional random walk in $a$ at nearly constant eccentricity $e$. 
The typical step-size of the random walk is related to the root-mean-square energy exchanged during flybys; 
we denote this quantity $\delta E_{p}$. 
After $N_{\rm flyby} \gg 1$ flybys, the fractional energy drift change $x$ may be regarded as a random variable with probability density given by 
\begin{equation} \label{eq:flyby_gaussian}
    f(x; \sigma_{p}, N_{\rm flyby}) = \frac{1}{\sqrt{2 \pi N_{\rm flyby}} \sigma_{p}} \exp\left( - \frac{x^{2}}{2N_{\rm flyby} \sigma_{p}^{2}} \right),
\end{equation}
where $\sigma_{p} \equiv \delta E_{p} / |E_{\rm i}|$ is the step size of the random walk in units of the scattering planet's initial energy. 


The typical step-size of the random walk can be estimated by geometric reasoning: 
If incoming planetesimals have a flat distribution of pericenter distances $r_{\rm p}$ and typical inclinations of at least a few degrees, 
and if gravitational focusing is negligible, 
then the probability of a flyby within a distance $s$ is proportional to $s^{2}$ 
\citep[e.g.,][]{Li+2022, Torres+2023}. 
Meanwhile, the energy scale of each encounter is $\sim G m m_{p}/s$. 
Thus, the typical energy exchange per flyby scales like $s^{2} \times (1/s) \propto s$: 
flybys with large $s$ contribute more to the planet's random walk in energy on average. 
On the other hand, the effects of flybys with $s \gg R_{\rm H}$ are drowned out by planet--planet interactions. 
Consequently, the cumulative effect of many flybys is dominated by those with $s \sim R_{\rm H}$. 
Setting $\delta E_{p} = G m m_{p} / R_{\rm H}$ and $E_{\rm i} = - G m m_{p} / 2 a$, we obtain
\begin{subequations} \label{eq:sigma-single-mass}
\begin{align}
    \sigma_{p} &= \frac{m_{p}}{2 M} \left( \frac{3 M}{m} \right)^{1/3} \\
    &= 9 \times 10^{-5} \left( \frac{m_{p}}{\ME} \right) \left( \frac{M}{\MSol} \right)^{-2/3} \left( \frac{m}{5 \ME} \right)^{-1/3}.
\end{align} 
This estimate is in good agreement with our numerical results. 
\end{subequations}

So far, we have only considered planetesimals of a single mass. 
It is straightforward to generalize equations (\ref{eq:sigma-single-mass}) to planetesimals drawn from an arbitrary mass function $\dif \mathcal{N}_{p} / \dif m_{p}$ (see Appendix \ref{appendix:sigma-multimass}). 
The distribution of $x$ after $N_{\rm flyby}$ flybys is then given by equation (\ref{eq:flyby_gaussian}), 
where we replace $\sigma_{p}$ with an effective value $\sigma$ 
given by a certain integral over the mass function (equation \ref{eq:sigma-multi-mass}). 
For a power-law distribution $\dif \mathcal{N}_{p} / \dif m_{p} \propto m_{p}^{-\alpha}$ ($1 < \alpha < 2$) 
over $m_{\rm min} \leq m_{p} \leq m_{\rm max}$, 
it can be shown that
\begin{subequations} \label{eq:sigma-powerlaw-mass}
\begin{align}
    \sigma^{2} &= \frac{2-\alpha}{3-\alpha} \frac{m_{\rm tot} m_{\rm max}}{M^{2}} \left( \frac{M}{m} \right)^{2/3} \left[ 1 - \left( \frac{m_{\rm min}}{m_{\rm max}} \right)^{3-\alpha} \right] \\
    &\simeq \frac{2-\alpha}{3-\alpha} \frac{m_{\rm tot} m_{\rm max}}{M^{2}} \left( \frac{M}{m} \right)^{2/3} ,
\end{align}
where $m_{\rm tot}$ is the combined mass of all incoming planetesimals. 
We take the limit $m_{\rm min} \ll m_{\rm max}$ on the second line. 
The most massive planetesimals dominate the cumulative scattering effect despite being relatively few in number. 
Observed planetesimal populations typically exhibit $1.5 \lesssim \alpha \lesssim 2$; 
for a population in collisional equilibrium, $\alpha = 11/6$ and $(2 - \alpha) / (3 - \alpha) = 1/7$ \citep{Dohnanyi1969}. 
Thus, a realistic planetesimal population has a somewhat less efficient cumulative scattering effect than a single-mass population. 
\end{subequations}


We now estimate the probability of escaping resonance through repeated flybys.  
Consider an ensemble of identical systems that undergo random transitions between adjacent level curves of $\mathcal{H}$ due to flybys, starting from the stable fixed point. 
Each level curve encloses an area $A \sim I_{\rm max}^{2}$ in phase space, 
and the area enclosed by the separatrix is $A_{\rm lib} \sim (\Delta I_{\rm lib})^{2}$. 
After many flybys, a typical system occupies a level curve with $I_{\rm max} \sim x \sim \sigma \sqrt{N_{\rm flyby}}$; 
the ensemble thus covers a characteristic area $A_{N} \sim N_{\rm flyby} \sigma^{2}$. 
According to the ergodic hypothesis, the probability that a particular system lies outside the separatrix after $N_{\rm flyby}$ encounters or fewer --
i.e. the fraction of systems having escaped resonance by this point -- is
\begin{subequations}
\begin{align}
    \mathcal{P}_{\rm esc} &\sim \frac{A_{N}}{A_{\rm lib}} \sim \frac{N_{\rm flyby} \sigma^{2}}{(\Delta I_{\rm lib})^{2}} \\
    &\sim 0.0025 \left( \frac{{\rm RCM}}{\ME} \right)^{2} \left( \frac{M}{\MSol} \right)^{-4/3} \nonumber \\
    & \hspace{1.5cm} \times \left( \frac{m}{5 \ME} \right)^{-2/3} \left( \frac{\Delta I_{\rm lib}}{10^{-3}} \right)^{-2},
\end{align} 
where 
\begin{equation}
    {\rm RCM} \equiv m_{p} \sqrt{N_{\rm flyby}}    
\end{equation}
is the ``root-cumulative'' mass encountered by the chain. 
For consistency with the experimental setup in Section \ref{sec:scattering_expts}, we evaluated $\sigma$ here using equation (\ref{eq:sigma-single-mass}). 
This estimate agrees with our numerical results within an order of magnitude (see Figure \ref{fig:CDF_N_m}). 
\end{subequations}

\section{Scattering Experiments for Repeated Flybys} \label{sec:scattering_expts}

\subsection{Setup}
\label{sec:setup}

We perform numerical experiments to investigate the breaking-the-chain process triggered by planetesimal flybys.
This process corresponds to the evolution of the system from the ``Newborn'' to the ``Young'' stage in Figure~\ref{fig:cartoon}.

The initial condition consists of a central star with mass $M=M_{\Sun}$ and three planets $m_1=m_2=m_3=1.5\times10^{-5}M_{\Sun} \simeq 5M_{\Earth}$ in a MMR chain.
The inner-most planet is at $a_1=0.1$au, corresponding to an orbital period $P_1=0.03$ year. 
The outer two planets are placed accordingly so the ratios of their orbital periods are $P_2:P_1 \simeq 3:2$ and $P_3:P_2\simeq j:(j-1)$.
The planets are coplanar and have negligible eccentricities; the evolution of their phase terms is characterized by the libration of the resonant angles
\begin{align}
\label{eq:theta12}
    \theta_{\rm 12} & = 3\lambda_2 - 2\lambda_1 - \varpi_1, \\
\label{eq:theta23}
    \theta_{\rm 23} & = j\lambda_3 - (j-1)\lambda_2 - \varpi_2,    
\end{align}
where $\lambda_{i}$ and $\varpi_{i}$ are the mean longitude and the longitude of pericenter for planet $i$, respectively.\footnote{For first-order resonances between comparable-mass planets, there are two resonant angles for each pair to be tracked in principle,
which differ according to which apsidal angle $\varpi$ appears in equations (\ref{eq:theta12}, \ref{eq:theta23}). 
For brevity, we present results for the specific angles $\theta_{12}$ and $\theta_{23}$ given above. 
Using their counterparts instead would not significantly change our results.}
In this Letter, we present results for $j=4$ (i.e., $P_{3}:P_{2} \simeq 4:3$) as a fiducial case; 
we have found that other $j$ values yield qualitatively similar results.

We then perturb the system with flybys of planetesimals on highly eccentric orbits originating from the outer planetary system via some dynamical process(es). 
We study two multi-flyby scenarios:
\begin{itemize}
    \item \textbf{Independent flybys} of multiple planetesimals, injected one-at-a-time into the planetary system on parabolic orbits.
    \item \textbf{Recurring flybys} of a single planetesimal on a bound eccentric orbit, passing through the inner planetary system repeatedly.
\end{itemize}
In real exoplanetary systems, the flux of incoming planetesimals may be a mixture of both. 
We will show that the results of both cases are indistinguishable.

\begin{table}[th!]
\centering
\caption{Details on the parameters in the flyby simulations (Section \ref{sec:scattering_expts}).
The \textbf{upper table} shows the planetary orbital elements in the initial MMR chain (formed via Type I migration, before flybys).
The \textbf{lower table} lists parameters we used for the inbound perturbers, with $\mathcal{U}$ representing uniform distributions; the longitudes of pericenter and ascending nodes are drown randomly from $\mathcal{U}(0,2\pi)$.}
\label{tab:sims}
\vspace{-3mm}
\begin{tabular}{ccccccc}
\hline\hline
$P_3:P_2$ & $a_1$ [au] & $a_2$ [au] & $a_3$ [au] & $e_1$ & $e_2$ & $e_3$ \\
\hline
$\simeq4:3$ & 0.1006 & 0.1320 & 0.1599 & 0.012 & 0.019 &  0.010 \\
\hline
\end{tabular}
\\
\vspace{2mm}
\begin{tabular}{lc}
\hline\hline
Parameters &  Values (independent / recurring) \\
\hline
Mass   & $10^{-3}M_{\Earth}$ - $10M_{\Earth}$ \\ 
Semi-major axis   &  $\infty$ $\quad / \quad$ 3au  \\
Eccentricity\footnote{The initial planetesimal eccentricity is not a free parameter, but is determined by the semi-major axis and the pericenter distance.} & 1.0 $\quad / \quad$  0.95 - 0.997 \\
Pericenter   & $\mathcal{U}(0.01\text{au},0.15\text{au})$  \\
Inclination   & $\mathcal{U}(1^{\circ},5^{\circ})$ \\
Pericenter time\footnote{The flight for the perturbers to reach the pericenter.} & $\mathcal{U}(0.5\text{yr}, 0.86\text{yr})$ \\
\hline
\end{tabular}
\end{table}

We carry out our numerical experiments using the $N$-body code {\tt REBOUND}~\citep{Rein.2012.A&A}.
The resonance chain is set up through the Type I migration scheme~\citep[based on][]{Cresswell.2008.A&A,Pichierri.2018.CeMDA,Kajtazi.2023.A&A} implemented in {\tt REBOUNDx}~\citep{Tamayo.2020.MNRAS}.
Details on the properties of the resulting planetary orbital elements can be found in Table~\ref{tab:sims}.

After the MMR chains are formed, we take the resulting systems as the initial condition for our flyby experiments. 
For independent flybys, we inject $N_{\rm flyby}$ planetesimals to the planetary system at a rate of 1 per 1.5 years.
In each experiment, the planetesimals are assumed to all have mass $m_{\rm p}$, while their incoming trajectories are random and independent.
For recurring flybys, we inject a planetesimal of mass $m_{\rm p}$ and simulate it passage through the planetary system for $N_{\rm flyby}$ times. 
After each passage, we use its outbound orbits to predict its return trajectory;
rather than simulating the complete orbit, we teleport the outbound planetesimals to its return trajectory to reduce the integration time.
This teleport-to-reenter procedure is done also at a rate of once per 1.5 years.
The flyby processes are simulated with the {\tt IAS15} integrator~\citep{Rein.2015.MNRAS}.
Details on the planetesimals' incoming trajectories can be found in Table~\ref{tab:sims}.

Finally, after all flybys complete, the planetary disk is further integrated for another $50$ years, allowing us to monitor the resulting behaviors of the resonance angles.

To quantitatively assess the net effect of planetesimal flybys, we perform a suite of experiment by varying the values of $N_{\rm flyby}$ and $m_{\rm p}$, and run a large ensemble of simulations for each $(m_{\rm p}, N_{\rm flyby})$ combination.

\begin{figure*}[t]
    \centering
    \includegraphics[width=0.95\textwidth]{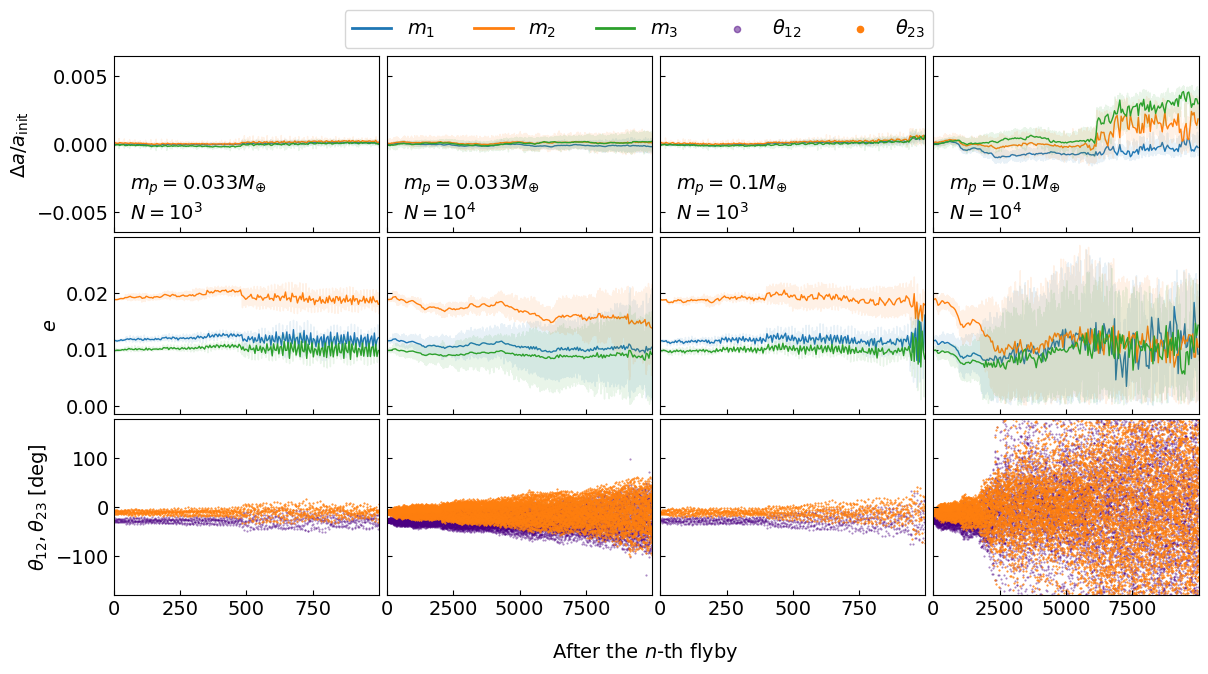}
    \caption{Example evolution of systems with 3:2-4:3 MMR chain perturbed by independent planetesimal flybys.
    Each systems receive $N$ flybys by planetesimals with mass $m_{\rm p}$ at a rate of 1 flyby per 1.5 years. 
    Each column shows the result with different $N$ and $m_{\rm p}$.
    \textbf{Top:} time evolution for changes of planetary semi-major axes ($\Delta a = a-a_{\rm init}$). 
    The faint lines show the measurement of $\Delta a/a$ after each flyby, while the solid curves show the average values from 200 uniformly spaced time bins.
    \textbf{Middle:} same as the top row, except showing the evolution of planetary eccentricities.   \textbf{Bottom:} measurement of the MMR angles $\theta_{12}$ and $\theta_{23}$, defined as in Equations~\eqref{eq:theta12} and~\eqref{eq:theta23}, after each flyby.
    }
    \label{fig:theta_examples_aet}
\end{figure*}


\subsection{Example Evolutions and Outcome Categorization}
\label{sec:result-Example}

We first illustrate how repeated flybys perturb the resonance chain. 
Figure~\ref{fig:theta_examples_aet} shows four example evolutions of the 3:2-4:3 resonant chains during a sequence of independent planetesimal flybys.
The columns correspond to different combinations of $(m_{\rm p}, N_{\rm flyby})$.

The upper panels show the evolution of the planetary semi-major axes $a$'s.
Each flyby induces a small drift in $a$, together with a few strong encounters producing larger $a$ shifts. 
The net effects of all the flybys slowly accumulate and scale with both $N_{\rm flyby}$ and $m_{\rm p}$. 
The middle panels show the evolution of the eccentricity $e$'s.  
While the time-average of the eccentricities evolves slowly (solid lines), their instantaneous values undergo rapid, large amplitude oscillations (faint lines).
Both the $a$ and $e$ evolutions are consistent with the anticipated behavior from the random-walk picture described in Section~\ref{sec:analytical-model}.

The bottom panels of Figure~\ref{fig:theta_examples_aet} track the values of the resonant angles $\theta_{12}$ and $\theta_{23}$ after each flyby.
Initially, both angles librate tightly around their stable fixed points (in this case, $0^{\circ}$), so the measurements are concentrated around it.
As the cumulative perturbations grow, measured values of $\theta_{12}$ and $\theta_{23}$ begin to spread, suggesting a gradual disruption of the resonances.

\begin{figure}[th]
    \centering
    \includegraphics[width=0.48\textwidth]{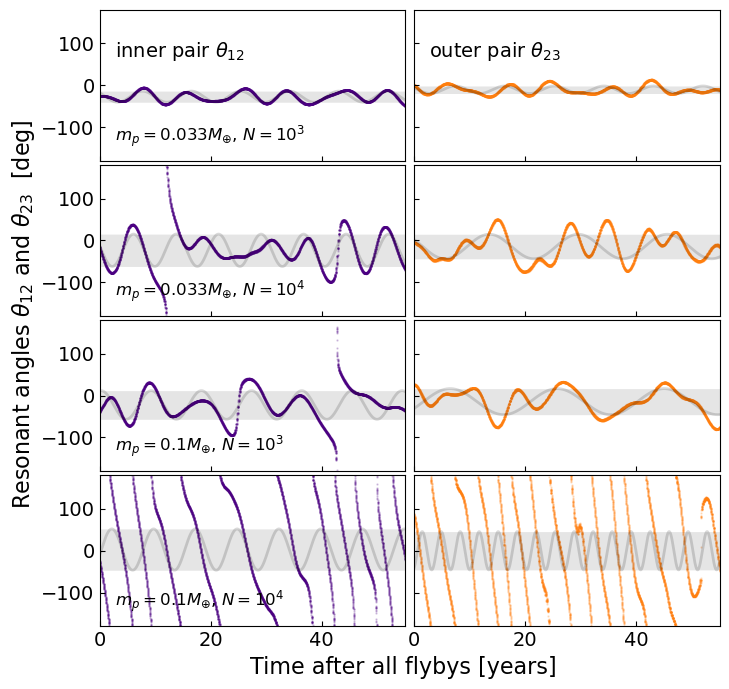}
    \caption{Time evolution of $\theta_{12}$ and $\theta_{23}$ in the 50 years after the last flyby is over, from the same simulations as in Figure~\ref{fig:theta_examples_aet}. 
    The gray lines show the sinusoidal fit to the time evolution (Equations~\ref{eq:theta12_sin} and~\ref{eq:theta23_sin}), while the shade regions represent the ranges of the sine curves.
    }
    \label{fig:theta_examples_thetas}
\end{figure}

Figure~\ref{fig:theta_examples_thetas} shows the resulting $50$-year evolutions of $\theta_{12}$ and $\theta_{23}$ after all flybys have completed.
The final behaviors of $\theta_{12}$ and $\theta_{23}$ fall into three categories:
\begin{itemize}
    \item \textbf{No breaking (top-row panels):} both resonant angles $\theta_{12}$ and $\theta_{23}$ remain librating after the flybys, even though their libration amplitudes have been perturbed to increase.
    \item \textbf{Partial breaking (middle-two-row panels):} the planetary system evolves in a mixed-mode where one or both resonant angles intermittently circulate. 
    Although both angles still mostly librate, these resonances are at the onset of being disrupted.
    \item \textbf{Complete breaking (bottom-row panels):} one or both of $\theta_{12}$ and $\theta_{23}$ circulate continuously, indicating the original MMRs have been disrupted.    
\end{itemize}
From no breakings to complete breakings, increasingly stronger cumulative perturbations are required.
Similar evolutions are also obtained in our experiments for the recurring-flyby scenario.

\subsection{Breaking Probability}

To quantify the probability of perturbing and breaking the resonances, we now analyze the full ensembles of the flyby simulations.
For each run, we fit the resulting $50$-year evolution of the resonant angles $\theta_{12}$ and $\theta_{23}$ to a sinusoidal ansatz,
\begin{align}
\label{eq:theta12_sin}
    \theta_{12} & \approx A_{12}\sin\left(\omega_{12} t + \phi_{12} \right) + C_{12}, \\
\label{eq:theta23_sin}
    \theta_{23} & \approx A_{23}\sin\left(\omega_{23} t + \phi_{23} \right) + C_{23};
\end{align}
and adopt the best fit values for $A_{12}$ and $A_{23}$ as the final libration amplitudes; we then use the larger of the two, $A\equiv\max(A_{12},A_{23})$, as a measure of the overall libration amplitude of the resonance chain.

We also calculate the wrap numbers, 
\begin{align}
    K_{12} & = \left|\Delta \theta_{12,{\rm unwrap}}\right| / 360^{\circ}\\
    K_{23} & = \left|\Delta \theta_{23,{\rm unwrap}}\right| / 360^{\circ},
\end{align}
with $\Delta \theta_{12,{\rm unwrap}}$ and $\Delta \theta_{23,{\rm unwrap}}$ being the net change of the unwrapped $\theta_{12}$ and $\theta_{23}$; 
$K_{12}$ and $K_{23}$ indicate how many full cycles of circulation each resonance angle completes during the 50-year follow-up integrations.

A system is categorized as completely breaking from the MMR chain if either $K_{12}$ or $K_{23}>10$, partially breaking if either $K_{12}$ or $K_{23}$ is greater than 1, and no breaking if otherwise.
While the majority of our simulated systems remain unbroken, we further distinguish the outcomes by their measured libration amplitude $A>20^{\circ}$ and $A>10^{\circ}$.

\begin{figure}[th!]
    \centering
    \includegraphics[width=0.45\textwidth]{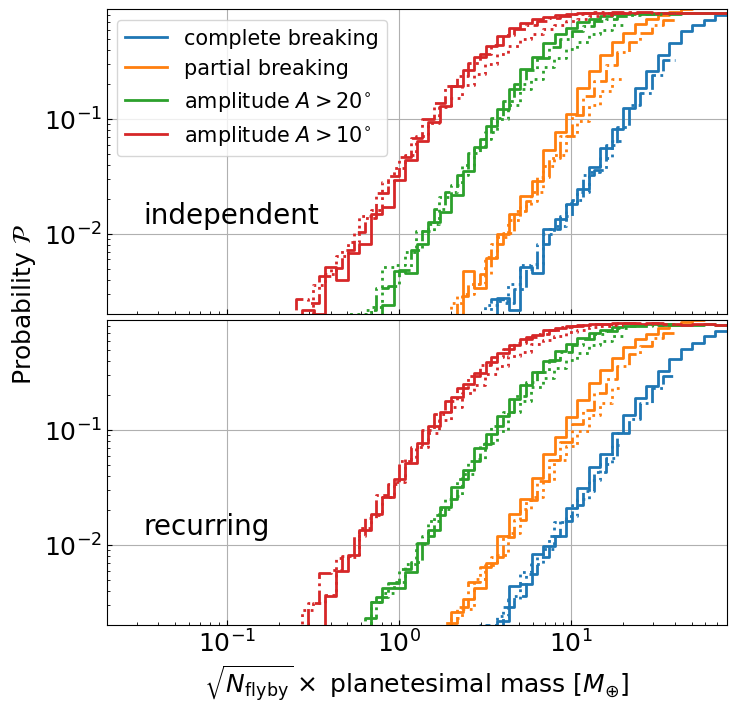}
    \caption{Probability of different resonance perturbation outcomes as a function of the root-cumulative perturber mass, ${\rm RCM} \equiv m_p\sqrt{N_{\rm flyby}}$ with $m_p$ as the mass of the planetesimal.
    The upper panel shows systems perturbed by independent flybys, while the lower panel shows results for recurring flybys of a single planetesimal.
    The solid, dashed-dotted, and dotted lines are the results from simulations with $N_{\rm flybys}=64$, $16$, and $4$, respectively. 
}
    \label{fig:CDF_N_m}
\end{figure}

Figure~\ref{fig:CDF_N_m} shows the probability for each outcome as a function of ${\rm RCM} = m_p\sqrt{N_{\rm flyby}}$.
Two clear trends emerge: (1) despite different $N_{\rm flyby}$ and $m_p$ in the simulations, the probability $\mathcal{P}$ of getting each outcome depends only on the RCM; (2) the probabilities $\mathcal{P}$'s are the same for both the independent and recurring flyby scenarios.

The probabilities follow the relation $\mathcal{P} \propto {\rm RCM}^2$ when they are small. 
Curve fittings find that, for $\mathcal{P} < 0.3$,
\begin{align}
\label{eq:P-fitting-cb}
    \mathcal{P}_{\rm complete} & = 2.1 \times10^{-4} \left(\frac{m_p}{M_{\oplus}}\right)^2N_{\rm flyby}, \\
\label{eq:P-fitting-pb}
    \mathcal{P}_{\rm partial} & = 7.4 \times10^{-4} \left(\frac{m_p}{M_{\oplus}}\right)^2N_{\rm flyby}, \\
\label{eq:P-fitting-A20}
    \mathcal{P}_{\rm A20} & = 5.7 \times10^{-3} \left(\frac{m_p}{M_{\oplus}}\right)^2N_{\rm flyby}, \\
\label{eq:P-fitting-pb}
    \mathcal{P}_{\rm A10} & = 3.2 \times10^{-2} \left(\frac{m_p}{M_{\oplus}}\right)^2N_{\rm flyby},
\end{align}
for getting complete breaking, partial breaking, libration with $A>20^{\circ}$, and libration with $A>10^{\circ}$, respectively. 
These power-law probability functions break down at $\mathcal{P}\gtrsim0.3$ as they saturate. 


\subsection{Implication}
\label{sec:fiducial-simulation-implication}

One important implication of the probability being dependent solely on the RCM is that, even when the planetesimals are low-mass, they may still break the resonance chain through a large enough number of flybys.

It should be noted that, in the recurring scenario, a single planetesimal can perform multiple flybys.
Without the other perturbers, a bound planetesimal would repeatedly fly through the inner planetary systems until it collides with a planet or is dynamically ejected from the planetary system.
The probability of collision per flyby can be estimated as $p_{\rm coll} \sim R_{\rm P}^2/a_{\rm P}^2$,
where $R_{\rm P}$ and $a_{\rm P}$ are the physical and the orbital radii of the planets, respectively.
Hence, the number of flybys the planetesimal can perform before collision is roughly
\begin{align}
    N_{\rm coll} \sim \frac{1}{p_{\rm coll}} \sim \frac{a_{\rm P}^2}{R_{\rm P}^2}. 
\end{align}
The required number of flybys to eject a planetesimal can be estimated by
\begin{align}
    \frac{GM_{*}}{2a_{\rm p}} \sim \sqrt{N_{\rm ejec}} \frac{G m_{\rm P}}{a_{\rm P}}
\end{align}
based on the random-kick diffusion of the planetesimal's orbital energy~\citep[e.g.][]{PuLai2021_cjscattering}. 
Using representative parameter values $a_{\rm P}=0.15$ au, $R_{\rm P}=2R_{\Earth}$, $m_{\rm P}=5M_{\Earth}$, and $a_{\rm p} = 3$ au, we have
\begin{align}
    N_{\rm coll} \sim N_{\rm ejec} \sim 3 \times 10^{6}.  
\end{align}
Therefore, the maximum number of flybys a single planetesimal can perform is possibly a few million.

\begin{figure}[t!]
    \centering
    \includegraphics[width=0.45\textwidth]{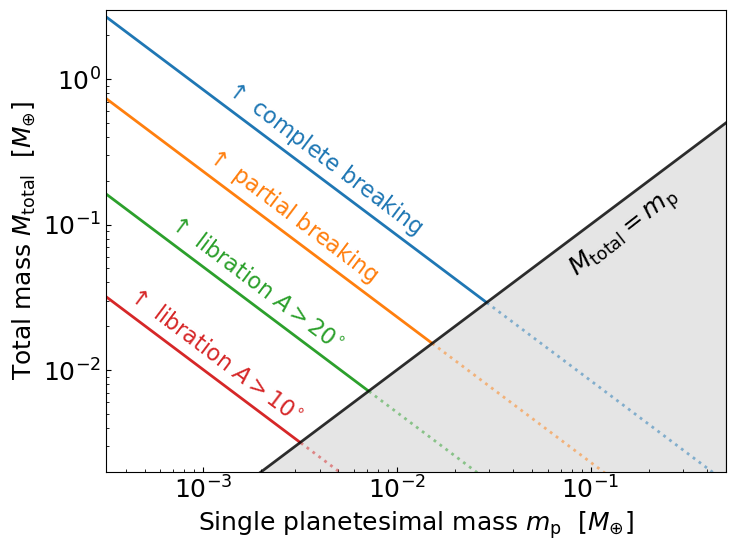}
    \caption{Critical total injected planetesimal mass $M_{\rm total} = m_{\rm p} N_{\rm object}$ required to perturb the resonant chain. 
    When $N_{\rm object}$ planetesimals of individual mass $m_{\rm p}$ are injected into the inner planetary system, the MMR chain breaks if $(m_{\rm p},M_{\rm total})$ lies above the blue line.
    Similar thresholds for other outcomes are shown by different colors.
    The shaded region corresponds to the unphysical regime where the $m_{\rm p}>M_{\rm total}$.
    }
    \label{fig:Nflyby-vs-m}
\end{figure}

Assuming each planetesimal can complete $\tilde{N} = 3\times10^6$ flybys before it is lost, how many distinct planetesimals (not the number of flybys) are needed to break the resonance chain?
For a given planetesimal mass $m_{\rm p}$, the required number of flybys for each level of perturbation can be estimated as $({\rm RCM_{50\%}}/m_{\rm p})^2$, where ${\rm RCM_{50\%}}$ is the threshold RCM for the corresponding probability to reach $\mathcal{P}=50\%$.
Hence, the total number of distinct planetesimals needed to be injected into the inner planetary system is
\begin{align}
    N_{\rm object} = \max\left[\frac{1}{\tilde{N}}\left(\frac{\rm RCM_{50\%}}{m_{\rm p}}\right)^2,1\right].
\end{align}
The total mass of these planetesimals would be $M_{\rm total} = m_{\rm p} N_{\rm object}$. 
We discuss how to relate this quantity to observations in Section \ref{sec:discussion}.


As shown in Figure~\ref{fig:CDF_N_m}, the threshold RCMs for a $50\%$ probability are
\begin{align}
    {\rm RCM_{50\%}} 
    & = \left\{51 M_{\Earth}, \ 26 M_{\Earth}, \ 12 M_{\Earth}, \ 6 M_{\Earth}\right\},
\end{align}
for complete breaking, partial breaking, libration with $A>20^{\circ}$ and $A>10^{\circ}$, respectively.
Based on these values, Figure~\ref{fig:Nflyby-vs-m} shows the estimated total mass of equal-mass planetesimals that must be injected into the planetary systems to produce each outcome.

\section{Instabilities in broken chains} \label{sec:instability}

After planets break out of resonance through flybys, how quickly do dynamical instabilities develop? 
It is critical to address this question in order to ensure consistency  
with the disappearance of MMRs among adolescent and mature planetary systems. 
To that end, we carry out long-term integrations of a subset of the systems studied in Section \ref{sec:scattering_expts}. 

In this discussion, we express time in units of the orbital period of the innermost planet in the system, denoted $P_{1}$, 
as this sets a practical limit on the length of our long-term integrations. 
For a representative $P_{1} \sim 10 \, {\rm d}$, $100 \Myr$ corresponds to $\sim 3 \times 10^{10}$ orbits. 
Our integrations cover only a fraction of this time. 

We select six sets of $100$ systems from the flyby simulations presented in Section \ref{sec:scattering_expts}. 
The RCMs for these sets are $m_{p} \sqrt{N_{\rm flyby}} \in \{ 11, 16, 22, 33, 47, 66 \} \ME$. 
Starting from snapshots immediately after the final planetesimal flyby, 
we integrate each system forward for $10^{9} P_{1}$ or until the onset of strong planet--planet scattering, whichever occurred first; 
for these integrations, we use the time-reversible hybrid integrator {\tt TRACE} \citep{Lu+2024_trace}. 
We detect strong scattering using {\tt REBOUND}'s built-in collision detection feature, 
assigning each planet a fictitious collision radius equal to half of its initial Hill radius. 
Upon the first Hill-sphere collision, we halt the simulation and recorded the elapsed time as the system's instability timescale $t_{\rm inst}$. 
We do not follow our simulations through the ensuing scattering phase, 
since the fates of unstable compact systems have been studied in detail elsewhere \citep[e.g.,][]{Hwang+2017, Izidoro2017_breakingchains, Izidoro2021_breakingchains, LiRixin2025_brokenchains}. 
We expect unstable systems to undergo giant impacts eventually ending in mergers, 
with the end-states roughly matching the mature KM population demographically. 

\begin{table}
\centering
\caption{Summary of long-term integration results. 
From left to right, the columns are the root-cumulative mass for each set of systems, 
the fraction $\mathcal{P}_{\rm inst}$ of systems with $t_{\rm inst} \leq 10^{9} P_{1}$,  
and the minimum and median $t_{\rm inst}$ of the unstable systems.}
\label{tab:instability}
\begin{tabular}{c|ccc}
\hline\hline
RCM [$\ME$] & $\mathcal{P}_{\rm inst}$ & Min.\ $t_{\rm inst}$ [$P_{1}$] & Med.\ $t_{\rm inst}$ [$P_{1}$] \\
\hline
11 & 0/100 & -- & -- \\ 
16 & 0/100 & -- & -- \\ 
22 & 0/100 & -- & -- \\ 
33 & 3/100 & $4.5 \times 10^{6}$ & $1.7 \times 10^{7}$ \\ 
47 & 4/100 & $1.0 \times 10^{6}$ & $5.5 \times 10^{7}$ \\ 
66 & 22/100 & $1.3 \times 10^{5}$ & $1.3 \times 10^{8}$ \\
\hline
\end{tabular}
\end{table}

We summarize the results of the extended integrations in Table \ref{tab:instability}. 
For each RCM, we report the fraction of systems rendered unstable within $10^{9} P_{1}$ and, 
where applicable, the minimum and median $t_{\rm inst}$ value for the unstable systems. 
Given the relatively small number of integrations in each set, 
these statistics have mainly illustrative value. 

None of the systems with ${\rm RCM} \leq 22 \ME$ went unstable within the allotted time. 
However, some instabilities occur for larger values: 
the instability fraction increases from $3\%$ at ${\rm RCM} = 33 \ME$ to $22\%$ at $66 \ME$. 
The minimum and median time to instability are $\sim 10^{5} \mbox{--} 10^{6} P_{1}$ and $\sim 10^{7} \mbox{--} 10^{8} P_{1}$, respectively. 

\begin{figure*}
    \centering
    \includegraphics[width=0.66\linewidth]{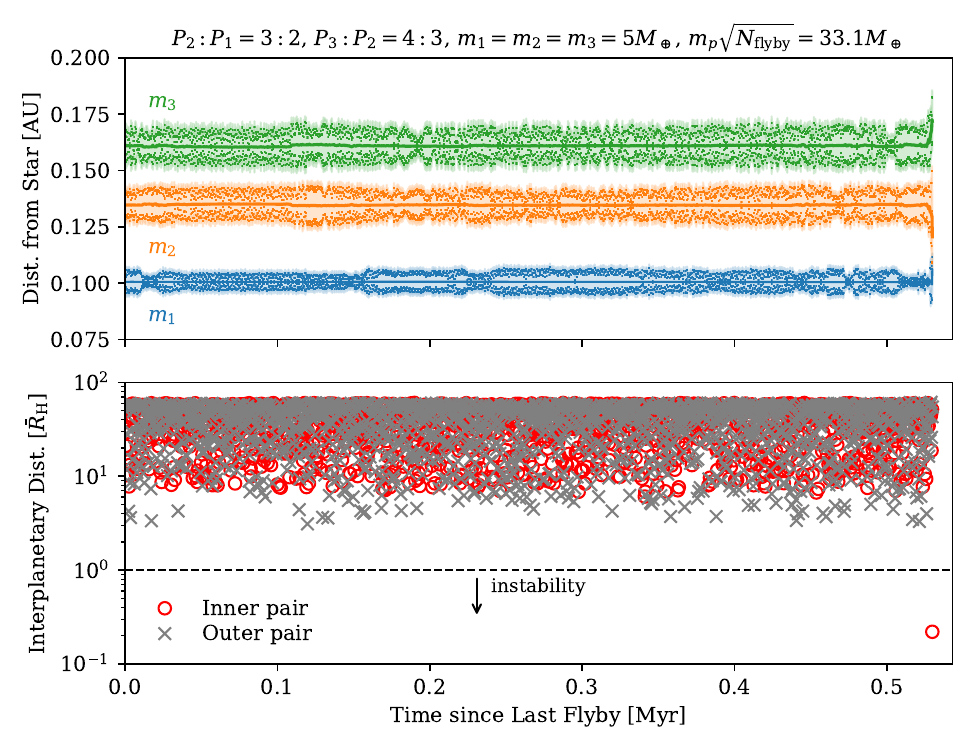}
    \caption{Onset of instability in a 3-planet system after a series of planetesimal flybys with $m_{p} \sqrt{N_{\rm flyby}} \approx 33 \ME$. 
    The planets were originally locked in a 3:2+4:3 resonant chain.
    Snapshots are shown at intervals of $10^{5} P_{1} \approx 3000 \yr$ and when the planetary Hill spheres first overlap. 
    {\bf Upper panel:} Orbital distances of each planet from the star over time: semi-major axis (solid curve) and periastron and apastron distance (small dots). 
    The shaded regions show the extent of each planet's Hill sphere inside and outside of its orbit. 
    {\bf Lower panel:} Three-dimensional separation between planets for the inner (red points) and outer (grey crosses) pairs, normalized by the mean Hill radius for each pair. 
    The horizontal line shows our threshold for strong scattering. 
    In this case, the inner pair crosses this threshold after $1.7 \times 10^{7} P_{1} \approx 0.53 \Myr$.}
    \label{fig:instability_example}
\end{figure*}

In Figure \ref{fig:instability_example}, we show the onset of instability in a single system after breaking out of resonance. 
This example was taken from the set with RCM $=33 \ME$. 
After an apparently quiescent stage lasting $1.7 \times 10^{7} P_{1}$, the planetary orbits begin to drift rapidly and increase in eccentricity, leading precipitately to orbit-crossing and the start of strong scattering. 
In this case, the inner planets experience the first close encounter; in some other runs, it is the outer pair. 

\begin{figure}
    \centering
    \includegraphics[width=\linewidth]{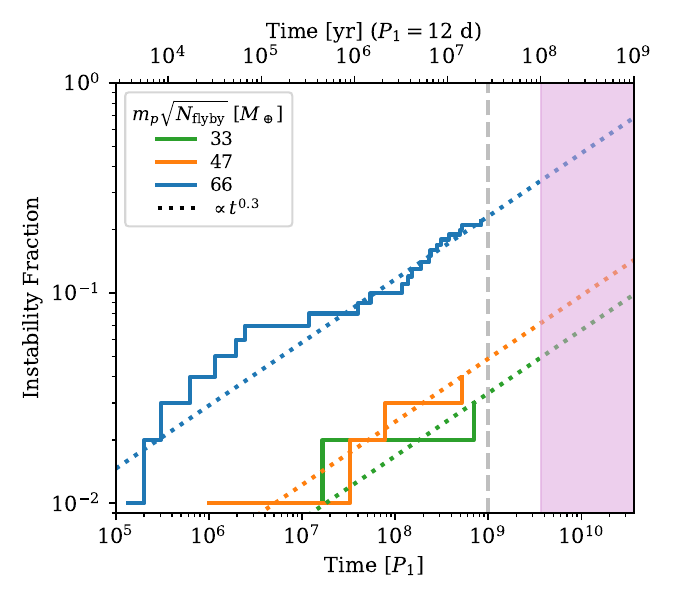}
    \caption{Instability fraction as a function of time in our extended integrations of planetary systems after planetesimal scattering for different RCM values (step plots). 
    The corresponding dotted lines show illustrative power-law extrapolations $\mathcal{P}_{\rm inst} \propto t^{0.3}$. 
    Time is given in units of $P_{1}$ on the lower horizontal axis
    and in years for $P_{1} = 12 \, {\rm d}$ on the upper. 
    The gray vertical line marks the maximum integration time. 
    The shaded region shows the age range of `adolescent' KM systems, 
    where a majority of MMR architectures disappear.}
    \label{fig:tinst_cdf}
\end{figure}

Again, our integrations cover only $\sim 3\%$ of the $100 \Myr$ instability timescale required by observations; 
hence, many apparently stable systems in our experiments may be unstable at late times. 
We gauge the potential for later instabilities by examining how the cumulative instability fraction scales with time. 
We show this in Figure \ref{fig:tinst_cdf} for our extended integrations with RCM $\geq 33 \ME$. 
To a very rough approximation, at early times the instability fraction $\mathcal{P}_{\rm inst}$ grows as a power law $\mathcal{P}_{\rm inst} \propto t^{\alpha}$ with $\alpha \sim 0.3$ 
\citep[cf.][]{Hwang+2017}. 
Using this fiducial power law scaling to extrapolate $\mathcal{P}_{\rm inst}$ forward from the latest recorded instability in each set, 
we estimate $\mathcal{P}_{\rm inst} = \{ 0.05, 0.07, 0.32 \}$ at $t = 100 \Myr$ and $\mathcal{P}_{\rm inst} = \{ 0.09, 0.13, 0.65 \}$ at $t = 1 \Gyr$. 
Crude though these estimates may be, it is encouraging that they approach the observed fraction of adolescent and mature KM systems {\it without} MMRs \citep{HamerSchlaufman2024_youngchains, Dai2024_chains}. 

\section{Discussion} \label{sec:discussion}

\subsection{Requirements to break the chains}

In Sections \ref{sec:scattering_expts}, we characterized the minimum mass budget of incoming objects required to break a resonant chain, 
using both numerical experiments and semi-analytical estimates. 
In order to break $> 50\%$ of systems out of resonance, we require a root-cumulative planetesimal mass $\gtrsim 50 \ME$. 
In Section \ref{sec:instability}, we demonstrated that dynamical instabilities develop 
within $10^{9}$ orbits in a reasonable fraction of these systems. 
Based on these findings, we conclude that repeated planetesimal flybys represent 
a plausible avenue for breaking the chains if the following conditions were met in a majority of KM systems: 
\begin{enumerate}
    \item A large reservoir of low-mass bodies, consisting of either planetesimals or small outer planets,  
    was present beyond the ice line after formation of the system, with a combined mass on par with the inner planets. 
    \item The outer regions of the system sustain sufficient dynamical activity to inject an appreciable fraction of that reservoir into the inner system on a timescale of $\sim 100 \Myr$ or less. 
\end{enumerate}
In this section, we discuss whether these conditions are compatible 
with current observational constraints on the architectures of planetary systems in general, and the KM systems in particular. 
Systems in which conditions (1) and (2) are not satisfied might correspond to the significant minority of mature KM systems where intricate MMR chains persist. 

\subsubsection{Presence of planetesimal reservoirs}

Many Sun-like stars possess planetesimal reservoirs left over from their formation. 
Debris disks beyond the ice line, which may be interpreted as Kuiper belt analogs, 
are detected via cold ($\lesssim 100  \, {\rm K}$) dust emission towards $\sim 20\%$ of mature FGK dwarfs \citep[e.g.,][]{Eiroa+2013_debrisdisks, Sibthorpe+2018_debrisdisks}. 
The occurrence fraction among young F and G dwarfs is higher, around $\sim 50\%$ \citep[e.g.,][]{Moor+2016}. 
Since current surveys are sensitive only to the brightest debris disks, 
these are almost certainly a lower bound on the actual occurrence rate. 
The optically thin dust masses in young debris disks are typically $\sim 1 \mbox{--} 100 \ME$ \citep[e.g.,][]{Michel+2021_dustmasses}, 
suggesting total solid mass budgets broadly compatible with our scenario. 

In principle, two further corrections should be applied to our estimated $M_{\rm total}$ to make a reasonable comparison with observations: 
We introduce a factor $F_{\alpha} \geq 1$ accounting for the planetesimal mass function (see Section \ref{sec:analytical-model}) 
and another factor $F_{\rm inj} \geq 1$ accounting for the efficiency of the dynamical injection mechanism acting on the reservoir. 
The predicted planetesimal reservoir mass is then $M_{\rm reservoir} = F_{\rm inj} F_{\alpha} M_{\rm total}$. 
Provided neither $F_{\alpha}$ nor $F_{\rm inj}$ is much greater than unity, the constraints on $M_{\rm total}$ derived from our single-mass scattering experiments (Figure \ref{fig:Nflyby-vs-m}) .

To give a concrete example, condition (1) appears to hold for AU~Mic, 
a young system ($22 \pm 3 \Myr$; \citealt{MamajekBell2014}) featuring both an inner (near-)MMR chain \citep{Plavchan+2020_AUMic, Martioli+2021_AUMic, Wittrock+2023_AUMic, Boldog+2025_AUMic} 
and an outer debris disk ($\sim 50 \mbox{--} 210 \AU$; \citealt{Kalas+2004_AUMicdisk}). 
If this system is a typical precursor to a mature KM system, 
then one might conclude that condition (1) is met in most cases. 

\subsubsection{Dynamical injection mechanism}

There are numerous dynamical channels by which to inject small bodies into the inner system.  
Many have been studied in detail in the context of metal pollution in white dwarfs, 
a widespread phenomenon believed to reflect the presence of dynamically active planetary systems. 
Here we briefly outline some mechanisms of potential interest:
\begin{itemize}
    \item {\it Outer planet secular interactions:} A significant fraction of the KM systems 
    are accompanied by outer planets ranging in size from super-Earths to super-Jupiters 
    \citep{ZhuWu2018, Bryan+2019, Herman+2019, BryanLee2024}. 
    Gravitational interactions between these planets and a planetesimal reservoir 
    could inject a steady stream of small bodies into the inner system during its early evolution \citep[e.g.,][]{Wisdom1983, Scholl1991, Malhotra1995, WuLithwick2011, LithwickWu2014, Smallwood+2018, Smallwood+2021, Li+2022_WDpollution, OConnor+2022}. 

    \item {\it Outer planet instabilities:} Many long-period exoplanets, especially gas giants, show signs of past dynamical instabilities. 
    Such instabilities would strongly perturb nearby planetesimal reservoirs, elevating the flux of small bodies through the inner planetary system \citep[e.g.,][]{DebesSigurdsson2002, Gomes+2005, FrewenHansen2014, Mustill+2018}. 
    The timing of such instabilities relative to the end of planet formation is highly sensitive to the system's initial conditions. 

    \item {\it Stellar companions:} Distant massive bodies, such as stellar or substellar companions to the host star, 
    can excite small bodies to extreme eccentricities via the von Zeipel--Lidov--Kozai effect \citep[e.g.,][]{Stephan+2017, PetrovichMunoz2017}. 
    This potentially relevant to KM-like systems formed in wide binaries with large inclinations relative to the planetesimal belt. 
    The companion's mass and orbit must be such that the injection occurs on a $\sim 100 \Myr$ timescale. 

    \item {\it External perturbations on an Oort-like cloud:} 
    If the planetesimal reservoir is located at an Oort-cloud-like distance ($\gtrsim 10^{3} \AU$), 
    then perturbations from the Galactic tidal field and passing stars may inject objects on parabolic orbits \citep[e.g.,][]{HeislerTremaine1986, OConnor+2023, PhamRein2024}. 
    However, in the solar neighborhood, the characteristic timescale for comet injection by these mechanisms is $\gtrsim 1 \Gyr$; 
    moreover, the fraction of injected bodies passing within $\sim 0.1 \AU$ of the central star would be small.

    \item {\it Self-gravity in a massive planetesimal belt:} 
    Some recent studies have advanced the idea that the collective self-gravity of planetesimals plays a significant role in the dynamical evolution of planetesimal belts \citep[e.g.,][]{Madigan+2018}. 
    We note in particular recent works suggesting that massive belts supporting a coherent eccentric mode 
    can drive a fraction of their constituent particles onto star-grazing orbits \citep[e.g.,][]{Akiba+2024}. 
    This possibility seems relevant in view of the large dust masses measured towards some young debris disks. 
    
\end{itemize}
A unifying aspect of the above is that the dynamical disturbances that break the chains in KM systems start at large distances and propagate inward. 
Thus, explaining the observed architectures of the KM systems may require that we widen our view of their dynamics in a literal sense. 

One distinguishing factor between the various possibilities listed above is the efficiency of planetesimal injection with high eccentricities. 
In systems where strong scattering dominates the dynamics beyond the ice line, the injection efficiency can be relatively small ($F_{\rm inj} \gg 1$) due to a statistical preference for ejections \citep[e.g.,][]{FrewenHansen2014, Mustill+2018}. 
On the other hand, distant or dynamically quiet perturbers, which exert their influence through secular interactions, tend to have relatively high injection efficiencies ($F_{\rm inj} \sim 1$; e.g., \citealt{PetrovichMunoz2017, OConnor+2022}). 

Finally, it is interesting and suggestive to note that both conditions (1) and (2) above were likely met in the early Solar System. 
The Jovian-planet region and early Kuiper Belt were littered with leftover planetesimals totaling $\sim 20 \mbox{--} 30 \ME$ \citep[e.g.,][]{Tsiganis+2005_Nice, Gomes+2005, Vokrouhlicky+2019_LPCs}. 
The orbits of the Jovian planets and surviving small bodies record a dynamical instability early in the system's lifetime, 
although its timing remains debated \citep[e.g.,][]{Tsiganis+2005_Nice, Clement+2018_gpinstability, deSousa+2020, Edwards+2024_gpinstability}. 
Although most of the planetesimals were ultimately ejected from the Solar System, 
many would have been injected into the terrestrial planet region as well. 
Thus, while the Solar System differs significantly from the KM systems, 
its reconstructed history provides conceptual support for our scenario.

\subsection{Caveats and extensions}


\subsubsection{Onset of instability}

How robust is our estimate of the required RCM value for timely instabilities in KM systems? 
We argue that, if anything, we have overestimated the threshold. 
Our reasoning is that some aspects of our simulations' setup may artificially inhibit instability. 
For example, we have studied equal-mass planetary systems only, 
which tend to have longer instability timescales than moderately unequal-mass systems. 
Similarly, the planets' initial eccentricities in our simulations are quite small, typically less than $\sim 0.02$. 
For comparison, the four planets in the Kepler-223 system's resonant chain  
have likely eccentricities of $\sim 0.05 \mbox{--} 0.15$ \citep{Mills2016_Kepler223chain}. 
Finally, systems with more than 3 planets have more degrees of freedom 
and hence may be more susceptible to chaotic diffusion after leaving resonance. 
Indeed, \citet{LiRixin2025_brokenchains} found a 6-planet chain could be destabilized by modestly increasing its libration amplitudes. 
These considerations reinforce our conclusion that RCM $\sim 30 \mbox{--} 60 \ME$ represents a conservative estimate of the disruption threshold for KM systems. 

\subsubsection{Dynamical effect of collisions}

In our analytical and numerical characterization of resonance escape, we considered purely gravitational interactions between planets and planetesimals. 
However, incoming planetesimals have a moderate probability of colliding with a planet over a long series of encounters (since $N_{\rm coll}\sim N_{\rm ejec}$, per Section \ref{sec:fiducial-simulation-implication}).
These collisions further facilitate the planets' escape from resonance \citep[see also][]{Raymond2022_trappist1planetesimals}. 
Here we quantify this additional contribution. 

Assuming that collisions between planets and planetesimals are well approximated as perfectly inelastic mergers, 
then each collision with a parabolic planetesimal of mass $m_{p}$ modifies the planet's orbital energy by 
\begin{equation}
    x_{\rm col} \sim \frac{m_{p}}{m} \sim \sigma_{p} \left( \frac{M}{m} \right)^{2/3}. 
\end{equation}
On average, each collision has a greater effect than each flyby by a factor of $(M/m)^{2/3} \sim 10^{3}$. 
However, each planetesimal can collide with a planet only once 
and takes a typical number of flybys $N_{\rm col} \sim 10^{6}$ to do so. 
Thus, the long-term effect of collisions with objects of mass $m_{p}$ 
can be taken into account in the random-walk model by increasing the variance per flyby 
by a fractional amount
\begin{subequations}
\begin{align}
    \left( \frac{\Delta (\sigma^{2})}{\sigma_{p}^{2}} \right)_{\rm col} &\sim \frac{\mathcal{P}_{\rm col}}{N_{\rm col}} \left( \frac{M}{m} \right)^{4/3} \\
    &\approx 0.5 \left( \frac{\mathcal{P}_{\rm col}}{0.1} \frac{10^{6}}{N_{\rm col}} \right) \left( \frac{M}{\MSol} \frac{5 \ME}{m} \right)^{4/3}.
\end{align}
The right-hand side in independent of $m_{p}$, 
so this correction is independent of the planetesimals' mass distribution. 
\end{subequations}

We conclude that planetesimal collisions increase the efficiency of resonance escape by $\sim 50\%$ compared to flybys alone. 
This slightly reduces the required planetesimal reservoir mass required to break the chains in KM systems. 

\subsubsection{Compatibility with other chain-breaking scenarios}

A planetesimal-driven scenario for breaking the chains is not mutually exclusive with other proposals in the literature. 
Different perturbative effects could conceivably dominate in different systems or even at different orbital scales within the same system. 
For example, the planetesimal flux per unit time generally falls off at short orbital periods; 
hence, tidal damping, photoevaporative mass loss, or secular spin--orbit resonances could have a greater effect in the most compact KM systems \citep{LithwickWu2012_repulsion, MatsumotoOgihara2020_breakingchains, Spalding2018_keplerobliquity, SpaldingMillholland2020_keplerobliquity, Faridani2023_sweeping, Faridani2025_sweeping}. 

\subsection{Summary and final remarks}

In this Letter, we have outlined a scenario for the breaking of resonant chains in compact exoplanet systems 
based on the scattering of planetesimals sourced from an outlying reservoir. 
Our main results are as follows:
\begin{enumerate}
    \item[(i)] Repeated planetesimal flybys can reliably disrupt resonant chains
    given a sufficiently large ``root-cumulative'' mass budget. 
    The required amount to disrupt resonances in $50 \%$ of KM-like systems is $m_{p} \sqrt{N_{\rm flyby}} \sim 3 \mbox{--} 30 \ME$, 
    depending on the disruption criterion.

    \item[(ii)] We characterize the process of resonance escape via flybys as a random walk in 
    the canonical phase space associated with MMRs between adjacent planets. 
    We derive a simple formula for the probability of escaping resonance as a function of the root-cumulative mass. 
    The result can be generalized to any MMR configuration and any mass function for the incoming planetesimals. 

    \item[(iii)] For $m_{p} \sqrt{N_{\rm flyby}} \gtrsim 30 \ME$, a significant fraction of simulated planetary systems are rendered unstable within $\sim 100 \Myr$, 
    consistent with the declining occurrence of MMRs with stellar age. 
    Owing to the limited time horizon of our $N$-body simulations, the actual minimum value for instabilities may be lower. 
\end{enumerate}

The main implications of our scenario for observations are as follows. First, young KM-like systems should be associated with massive ($\gtrsim 30 \ME$) outlying planetesimal reservoirs. 
Second, the non-resonant KM population may preferentially host dynamically active outer planets responsible for planetesimal injection at early times. 
Conversely, those KM systems where MMR chains have survived at late times would be expected to lack outer companions. 

If the breaking-the-chains hypothesis is correct, it would pose an interesting constraint on the origin of other, rarer exoplanet architectures. 
Ultra-short-period planets (USPs), small planets with orbital periods less than 1 day, are a prime example. 
Observed around $\sim 1\%$ of stars, usually as members of mature KM systems \citep[e.g.,][]{Schmidt+2024_USPMMR}, 
USPs likely form via planet--planet interactions on Gyr timescales \citep[e.g.,][]{Petrovich2019_USPs, PuLai2019_USPs, MillhollandSpalding2020} 
and might ultimately be destroyed in the same way \citep{OConnorLai2025}. 
A future study could examine USP formation and evolution within the breaking-the-chains paradigm 
in order to constrain the underlying initial conditions and dynamical mechanisms. 

\begin{acknowledgments}
    JL and CEO are supported by Center for Interdisciplinary Exploration and Research in Astrophysics (CIERA) Postdoctoral Fellowships. 
    This work used computing resources provided by Northwestern University and CIERA funded in part by NSF PHY-2406802. 
    This research was supported in part through the computational resources and staff contributions provided for the Quest high-performance computing facility at Northwestern University, 
    which is jointly supported by the Office of the Provost, the Office for Research, and Northwestern University Information Technology.
\end{acknowledgments}

\software{{\tt Matplotlib} \citep{Hunter2007_matplotlib}, {\tt NumPy} \citep{Harris+2020_NumPy}, {\tt REBOUND} \citep{Rein.2012.A&A}, {\tt REBOUNDx} \citep{Tamayo.2020.MNRAS}, {\tt SciPy} \citep{Virtanen2020_scipy}}

\appendix

\section{Planetesimal scattering with an arbitrary mass function} \label{appendix:sigma-multimass}

In Section \ref{sec:analytical-model}, we argued that, for encounters with planetesimals of a single mass $m_{p}$,  
the r.m.s.\ dimensionless energy exchange per flyby $\sigma_{p}$ is given by equation (\ref{eq:sigma-single-mass}). 
Here we derive an effective value $\sigma$ for encounters with planetesimals drawn from an arbitrary mass function. 

Let the incoming planetesimals consist of various species $p$, 
where $m_{p}$ and $\mathcal{N}_{p}$ are the individual mass and total number of planetesimals of each species; 
and let $\mathcal{N} \equiv \sum_{p} \mathcal{N}_{p}$ be the total planetesimal population. 
We postulate that the r.m.s.\ energy exchange per flyby $x$ 
is the average of the contributions from encounters with each incoming planetesimal, i.e.\ 
\begin{equation}
    x = \frac{1}{\mathcal{N}} \sum_{p} \sum_{q=1}^{\mathcal{N}_{p}} x_{p,q}, 
\end{equation}
where $x_{p,q}$ corresponds to the energy exchanged with the $q$th planetesimal of species $p$. 
Each $x_{p,q}$ is an independent Gaussian random variable with zero mean and variance $\sigma_{p}^{2}$. 
Hence, by the central limit theorem, $x$ is a Gaussian random variable with zero mean and variance $\sigma^{2} = (1/ \mathcal{N}) \sum_{p} \mathcal{N}_{p} \sigma_{p}^{2}$; 
or, in the continuum limit,
\begin{equation} \label{eq:sigma-multi-mass}
    \sigma^{2} = \frac{1}{\mathcal{N}} \int_{m_{\rm min}}^{m_{\rm max}} \sigma_{p}^{2}(m_{p}) \frac{\dif \mathcal{N}_{p}}{\dif m_{p}} \, \dif m_{p},
\end{equation}
where $\dif \mathcal{N}_{p} / \dif m_{p}$ represents the incremental mass-frequency distribution 
and where $m_{\rm min}$ ($m_{\rm max}$) is the minimum (maximum) planetesimal mass. 
After $N \gg 1$ flybys, the probability density of $x$ is given by equation (\ref{eq:flyby_gaussian}) with $\sigma$ substituted for $\sigma_{p}$. 
Equation (\ref{eq:sigma-powerlaw-mass}) gives the result for a power law distribution $\dif \mathcal{N}_{p} / \dif m_{p} \propto m_{p}^{-\alpha}$ 
with $1 < \alpha < 2$ and $m_{\rm min} \ll m_{\rm max}$.

\end{document}